\documentclass[proceedings]{JHEP}

\usepackage{amsfonts}
\usepackage{amssymb}
\usepackage{epsfig}
\usepackage{psfig}

\newcommand{\beqn}{\begin{eqnarray}}
\newcommand{\eeqn}{\end{eqnarray}}
\newcommand{\be}{\begin{equation}}
\newcommand{\ee}{\end{equation}}
\newcommand{\non}{\nonumber \\}
\newcommand{\tr}{\mbox{tr}} 
\newcommand{\Tr}{\mbox{Tr}}

\newcommand{\al}{\alpha^\prime}

\newcommand{\sz}{\scriptsize}

\preprint{hep-th/0002146} 
\title{A New Class of Supersymmetric Orientifolds with D-branes at Angles}
\conference{Quantum Aspects of Gauge Theories, Supersymmetry and Unification,
  Paris 1999}
\keywords{Superstring Vacua, D-branes}

\author{Ralph Blumenhagen\footnote{e-mail: blumenha@physik.hu-berlin.de}, 
Lars G\"orlich\footnote{e-mail: goerlich@physik.hu-berlin.de} and 
Boris K\"ors\footnote{e-mail: koers@physik.hu-berlin.de} \\
Humboldt Universit\"at zu Berlin, 
Institut f\"ur Physik, Invalidenstr. 110, 10115 Berlin, Germany}

\abstract{
We describe a new class of supersymmetric orientifolds which combine 
the world-sheet parity transformation $\Omega$ with 
a complex conjugation in the compact directions. 
As an example, we investigate in detail the orientifold of the $\mathbb{Z}_3$
toroidal orbifold in six and four dimensions.
We demonstrate how the solution
to the tadpole cancellation conditions, the resulting gauge groups 
and the massless spectra depend on
the choice of the complex structures on the tori, giving rise to a variety of
inequivalent models. We also summarize the results for the orientifolds of the
$\mathbb{Z}_4$, $\mathbb{Z}_6$, and
$\mathbb{Z}_6^\prime$ orbifolds in four and six dimensions.
}

\begin{document} 
\section{Introduction}

Orientifolds of type II string theories or, equivalently,  
compactifications of the type I open string theory 
have considerably enlarged our view on consistent
vacua of string theory.
The plethora of
Calabi-Yau compactifications known for the heterotic string carries over
to type I via the ten dimensional S-duality \cite{hepth9510169}. 
While in ten dimensions it involves an inversion of the string coupling, 
this is no longer
necessary in six and four dimensions. However, in the type I setting
one can perturbatively study backgrounds which  are non-perturbative
from the heterotic viewpoint, as they they involve lower dimensional
D-branes. Besides all those recent ideas arising in this context, like  
lower string scale unification, type I vacua still leave
a variety of open questions and puzzles.

First, employing the description of type I compactifications as 
orientifolds of
type II, one chooses an exactly solvable point in the moduli space 
of the Calabi-Yau manifold and performs the tadpole cancellation computation.
In all known cases these exactly solvable points are given  by toroidal 
orbifolds or Gepner models. In general it is not clear how to deform 
away from these special points in moduli space.
Second, at first sight surprisingly it appeared  that some of the 
four dimensional 
standard orbifolds do not allow cancellation of all massless tadpoles
\cite{Aldazabal:1998mr}.
This perturbative inconsistency has been conjectured to be cured
due to non-perturbative states in \cite{hepth9804092}. 
However, not all cases for which perturbative inconsistencies arise
have been resolved in this fashion.

Finally, the fate of
the discrete NS-NS antisymmetric B-field modulus in the blown-up version and
its translation to the heterotic side have not been explored in detail. 
Related to the presence of a B-field background is the
occurance of extra multiplicities for certain open string sectors, which are
required in order to achieve tadpole cancellation and anomaly freedom. \\

Some of these issues may possibly be tackled in a new class of 
orientifold models which we have
presented in \cite{hepth9908130, hepth9912204}\footnote{Some of the examples
  have been recovered in a completely different approach in
  \cite{hepth9912218}.} 
and which has also been
analyzed in \cite{hepth9911190}. They combine the world-sheet
parity transformation with a reflection ${\cal R}$ of half of the internal 
coordinates, which can be written  as
a complex conjugation on any two dimensional torus.
Note, that this combination still preserves half of the supersymmetry. 
The new features arising in this class of orientifolds are that the
operation $\Omega {\cal R}$ does not exchange twisted sectors and that the
internal B-field becomes a continuous modulus 
(whereas the off-diagonal entries 
of the metric get discretized).
The extra multiplicities for certain open string sectors 
arise very naturally as intersection numbers of D-branes. 
We find solutions for any of the orbifold groups we have
investigated. Even more surprisingly, we get more than just one orientifold per
orbifold group, distinguished by the particular choice of the complex
structures on the tori. From the  phenomenological point of view 
the four dimensional models are of little
interest, as they generically show  non-chiral matter content. \\

In section 2 we describe the general 
features of $\Omega{\cal R}$ orientifolds. In section
3 we investigate the $\mathbb{Z}_3$ example as the technically most simple
one in some detail. It already exhibits
all conceptually new aspects of $\Omega {\cal R}$ orientifolds.  
The section 4 finally summarizes results of our computations for
the $\mathbb{Z}_4$ and the two $\mathbb{Z}_6$ examples.

\section{$\Omega{\cal R}$ orientifolds}

The orientifolds with $\Omega {\cal R}$ projection combine the 
ordinary world sheet parity transformation $\Omega$ with a conjugation 
${\cal R}$ of the complex coordinates of the compactification torus $T^4$ or
$T^6$, for six- or four dimensional models respectively. The tori are taken to
factorize into two dimensional tori in all cases, with coordinates given by 
\beqn
X_i \equiv x_{10-2i}+i x_{11-2i},\quad i=1,2,3.
\eeqn
The action of ${\cal R}$ then reads 
\beqn
{\cal R}: X_i \mapsto \overline{X}_i.  
\eeqn
The complex structure of each two dimensional torus is further restricted by
the requirement of the orbifold group acting crystallographically. We
have chosen (up to rescaling) the $SU(3)$ or $SU(2)^2$ root lattices for
simplicity. The relative choice of
the complex structures on the various $T^2$ tori 
is of crucial importance for the gauge
group and the spectra, which leads to a variety of distinct models. 
Since ${\cal R}$ can also be considered as a reflection of one half of the 
compact real coordinates, $\Omega{\cal R}$ is a symmetry of 
type IIA in four and   
of type IIB in six dimensions. This operation is then accompanied by one 
of the well 
known cyclic $\mathbb{Z}_N$ orbifold groups preserving ${\cal N}=1$ 
supersymmetry. We denote
the groups by $\mathbb{Z}_N=\{ 1,\Theta, ...,\Theta^{N-1} \}$. The operation of
the generator is diagonal in the complex basis:
\beqn
\Theta : X_i \mapsto \exp \left( 2\pi i v_i \right) X_i
\eeqn  
and with opposite phases on the conjugate variables. 
Also the fermionic coordinates are being complexified in order to diagonalize
$\Theta$. The operation on the various ground states reads 
\beqn \label{Rongroundst}
{\cal R} : \vert s_0, s_1, s_2, s_3 \rangle & \mapsto & 
\vert s_0, -s_1, -s_2, -s_3 \rangle , \\ \nonumber 
\Theta :  \vert s_0, s_1, s_2, s_3 \rangle & \mapsto & 
 \exp \left( 2\pi i \vec v \cdot \vec s \right) 
\vert s_0, s_1, s_2, s_3 \rangle .
\eeqn
If in some twisted sector the ground  state is only a spinor of a subgroup of
$SO(8)$, the respective $s_i$ are set to zero formally. 
A further subtlety arises with the GSO projection which is ambiguous in the
twisted closed string and the open string sectors a priori. 
It can be fixed by imposing the $\Omega {\cal R}$ symmetry of the spectrum and
requiring supersymmetry. The entire orientifold group is given 
by $\mathbb{Z}_N \cup \Omega{\cal R} \mathbb{Z}_N$ and these models 
cannot be mapped by T-duality to standard orientifolds
$\mathbb{Z}_N \cup \Omega \mathbb{Z}_N$. Instead they are dual to  
asymmetric orientifolds $\hat\mathbb{Z}_N \cup \Omega 
\hat\mathbb{Z}_N$ with standard $\Omega$ projection. The way how the 
T-duality acts on the various moduli of the models has been investigated
in \cite{hepth9911190}. \\

Six dimensional $\mathbb{Z}_N$ orbifolds preserving  ${\cal N} =1$ 
supersymmetry are easily
defined by $v_1=1/N,$ $v_2=-1/N$. Orbifolds featuring ${\cal N} =1$ 
supersymmetry in four dimensions have been classified in 
\cite{DHVW1985,DHVW1986} 
and we display their action on the complex basis in terms of the 
$v_i$ in table \ref{tablezn}. 
\TABULAR[h]{|l|l|}{
\hline
$\mathbb{Z}_3 : v= (1,1,-2)/3$ & $\mathbb{Z}_8 : v= (1,3,-4)/8$ \\
$\mathbb{Z}_4 : v= (1,1,-2)/4$ & $\mathbb{Z}_8^\prime : v= (1,2,-3)/8$ \\
$\mathbb{Z}_6 : v= (1,1,-2)/6$ & $\mathbb{Z}_{12} : v= (1,4,-5)/12$ \\
$\mathbb{Z}^\prime_6 : v= (1,2,-3)/6$ & $\mathbb{Z}_{12}^\prime : 
v= (1,5,-6)/12$ \\
$\mathbb{Z}_7 : v= (1,2,-3)/7$ & \\
\hline}{
\label{tablezn}$\mathbb{Z}_N$ groups that preserve ${\cal N}=1$ in $d=4$}
We will discuss the $\mathbb{Z}_3$ example in greater detail and will
only briefly state the results
for the $\mathbb{Z}_4$, $\mathbb{Z}_6$ and $\mathbb{Z}_6^\prime$ cases.
For all the other orbifold groups we do neither expect any conceptual
novelties, nor will they produce more realistic models in terms of gauge groups
or massless spectra, while surely demanding a large amount of 
tedious computation. 
In contrast to our results for the $\Omega{\cal R}$ orientifold it has been
established for the ordinary four dimensional $\mathbb{Z}_4$ case that there
exists no solution to the tadpole condition. This is also the case for the
standard $\mathbb{Z}_8$, $\mathbb{Z}_8^\prime$ and $\mathbb{Z}_{12}$
orbifolds, which we believe to have solutions as $\Omega {\cal R}$
orientifolds, as well. \\

\section{The $\mathbb{Z}_3$ example}

In the following we describe the computations relevant for the $\mathbb{Z}_3$
orientifolds in $d=4$ and $d=6$ with $\Omega {\cal R}$ projection in
detail. The two kinds of models are technically very similar but a number of
subtle distinctions need to be made with respect to the dimensionality. 
We shall
mostly display the formulas for the six dimensional case and explain the
changes in four dimensions separately. The generator $\Theta$ of the orbifold
group acts diagonally by  
\beqn
\Theta : X_{1,2} \mapsto \exp \left( \pm2\pi i /3 \right) X_{1,2}
\eeqn
in six dimensions or by 
\beqn
\Theta : X_{1,2} & \mapsto & \exp \left( 2\pi i /3 \right) X_{1,2}, \non
\Theta : X_3 & \mapsto & \exp \left( -4\pi i /3 \right) X_3
\eeqn
in four dimensions. The torus we employ
is defined by the root lattice of $SU(3)$ with basis vectors of length
$\sqrt{2}$ for any of the complex directions,
being depicted in Figure \ref{figz3zellen}. It also displays the fixed points
of $\Theta$ by circles. 
\FIGURE[h]{
\makebox[6cm]{
 \epsfxsize=6cm
 \epsfysize=15cm
 \epsfbox{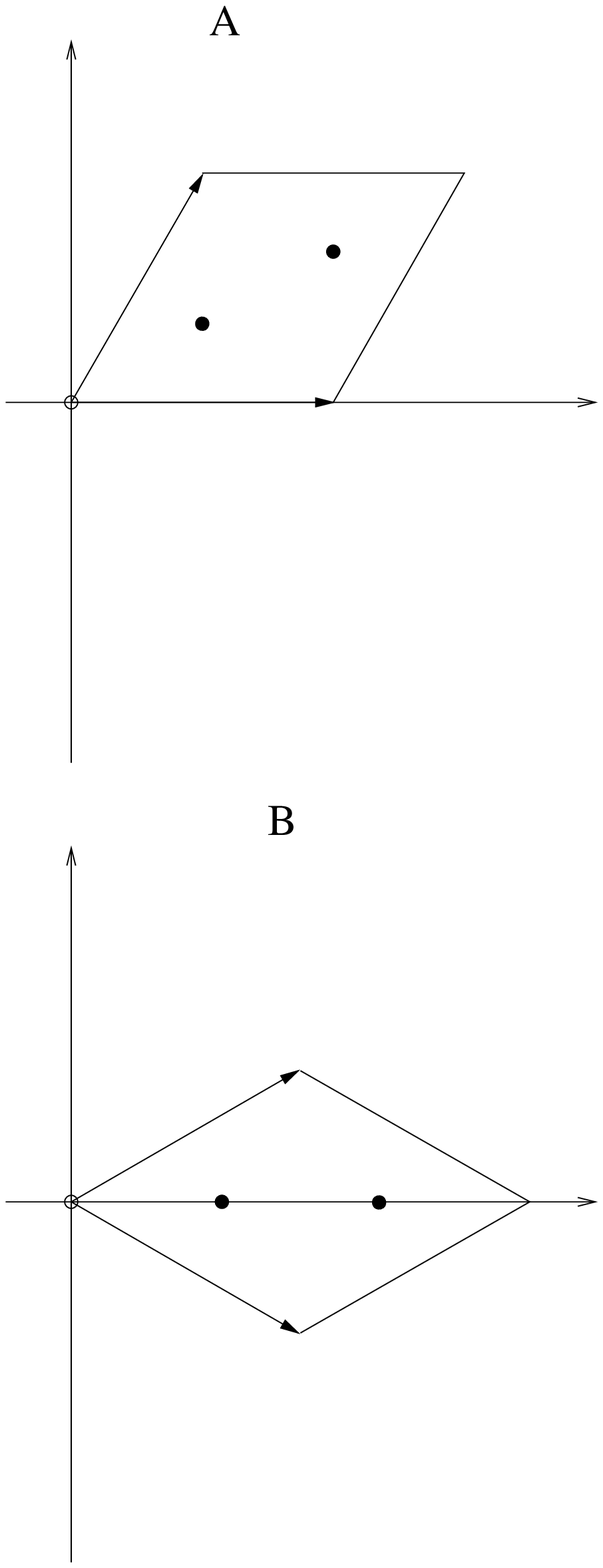}
}
\caption{\label{figz3zellen}The $\mathbb{Z}_3$ lattices}}
%
The action of ${\cal R}$ is chosen to be a reflection of the vertical axis,
leaving two inequivalent orientations of the lattice, denoted by {\bf A} and
{\bf B}. Both lattices allow a crystallographic reflection, but lead to
different results for the invariant
Kaluza-Klein (KK) momenta and winding (W) states 
as well as for the number of $\Omega {\cal R}$ 
invariant fixed points of $\Theta$. 

\subsection{Closed string sector: The Klein bottle}

In the closed string sector  we have twisted sectors for all powers 
of $\Theta$. The 
Fourier decomposition of the coordinate fields carries modings shifted by
$kv_i$ in the $k$-twisted sector. An important feature is that $\Omega$ as
well as ${\cal R}$ exchange twisted sectors implying that the twist fields are
left invariant by the combined action. We then get the $\Omega{\cal R}$
invariant oscillator excitations  
\beqn
\left( \Omega{\cal R}\right) \alpha_{n+kv_i} 
\tilde{\overline{\alpha}}_{n+kv_i} \left( \Omega{\cal R}\right)^{-1}  & = &
\alpha_{n+kv_i} \tilde{\overline{\alpha}}_{n+kv_i} , \non 
\left( \Omega{\cal R}\right) \overline{\alpha}_{n-kv_i} 
\tilde{\alpha}_{n-kv_i} \left( \Omega{\cal R}\right)^{-1}  & = &
\overline{\alpha}_{n-kv_i} \tilde{\alpha}_{n-kv_i} . \nonumber
\eeqn
The $\alpha$ denotes any kind of ladder operator in the 
$k$-twisted sector with $n\in \mathbb{Z}$ or $n\in \mathbb{Z}+1/2$.
The states are also invariant under the action of $\Theta$:
\beqn
\Theta \alpha_{n+kv_i} \tilde{\overline{\alpha}}_{n+kv_i} 
\Theta^{-1}  & = & 
\alpha_{n+kv_i} \tilde{\overline{\alpha}}_{n+kv_i}, \non
\Theta \overline{\alpha}_{n-kv_i} \tilde{\alpha}_{n-kv_i} 
\Theta^{-1}  & = & 
\overline{\alpha}_{n-kv_i} \tilde{\alpha}_{n-kv_i}. \nonumber
\eeqn
The left- and right-moving momenta are given by 
\beqn
p_{\rm L} &=& \frac{1}{i\sqrt{\al U_2 T_2}} \left( m_1 U - m_2 - \overline{T}
  \left( n_1 + U n_2 \right) \right) , \non \nonumber 
p_{\rm R} &=& \frac{1}{i\sqrt{\al U_2 T_2}} \left( m_1 U - m_2 - T
  \left( n_1 + U n_2 \right) \right) ,
\eeqn
where $U$ and $T$ are the complex structure and the complexified K\"ahler
modulus of the $T^2$. They are given in terms of the unit vectors spanning the
lattice:
\beqn
U = \frac{e_2}{e_1} &,& \quad T= iV, \\
e^{\rm {\bf A}}_1= \sqrt{2} &,& \quad e^{\rm {\bf A}}_2= \frac{1}{\sqrt{2}}+i 
  \sqrt{\frac{3}{2}} , \non
e^{\rm {\bf B}}_1= \sqrt{2}  &,& \quad 
e^{\rm {\bf B}}_2= \frac{1}{\sqrt{2}} +i \frac{1}{\sqrt{6}} , \nonumber
\eeqn 
where we have adopted the choice of Figure \ref{figz3zellen}. This allows to
determine the KK+W states invariant under $\Omega{\cal R} \Theta^k$. In fact,
the identity 
\beqn
\Theta^{-1/2} \left( \Omega{\cal R} 
\Theta^k \right) \Theta^{1/2} = \left( \Omega{\cal R} \Theta^k \right) \Theta 
\eeqn
and the $\Theta^{-1/2}$ symmetry of the lattice implies that their
contribution to the partition function is independent of $k$. \\

Using the above mentioned ingredients, the computation of 
the loop channel Klein bottle amplitude becomes a  straightforward exercise: 
\beqn
\label{kleinloop}
{\cal K} =& &  2^{d/2} c\int_0^\infty{\frac{dt}{t^{d/2+1}}\ \Tr_{\mbox{\sz U+T}} 
\left( \frac{\Omega{\cal R}}{2} \right. } \\ \nonumber
& &  
\left. \frac{\left( 1+\Theta + \Theta^2 \right)}{3} 
\frac{\left( 1+(-1)^F \right) }{2} e^{-2\pi t \left( L_0 + \bar{L}_0 \right) }
\right) .
\eeqn
We define $c \equiv V_d/ \left( 8\pi^2 \al \right)^{d/2}$ 
and the momentum integration in the non-compact space-time has already 
been performed.
While in the loop channel there are contributions in all twisted sectors, the
relation  
\beqn
\Omega{\cal R}\Theta^k = \Theta^{-k}\Omega{\cal R}
\eeqn
implies that only untwisted  closed string states  
propagate in the tree channel Klein bottle amplitude. 
More precisely, the modular transformation maps
the contributions of the $k$-twisted sector to the untwisted sector with the
term $\Theta^k$ of the orbifold projector $(1+\Theta+\cdots +\Theta^{N-1})/N$ inserted. 
In order for the corresponding boundary states to be invariant under $\Theta$,
it is necessary that the modular transformation produces the complete
projector in the tree channel. It turns out that the latter condition 
is not always met automatically. 
We consider the completion of the projector as the guiding principle, 
which determines  how to choose the relative
orientations of the $T^2$ lattices, say of {\bf A} or {\bf B} type. 
Whenever this loop-tree channel equivalence  is
satisfied we find consistent models. In case it is not, one is actually
missing states in the tree channel to complete the projector perturbatively, a
circumstance that may hint towards the existence of non-perturbative states,
that cure the discrepancy \cite{hepth9904007}. \\

Luckily, for the $\mathbb{Z}_3$ orbifold this condition does not 
impose any constraints, while it does so for all the other orbifold
groups. Thus we have three inequivalent choices of the lattice (their complex
structures) in six dimensions: 
\beqn
{\bf AA},{\bf AB},{\bf BB}
\eeqn 
and four in four dimensions: 
\beqn
{\bf AAA},{\bf AAB},{\bf ABB},{\bf BBB}.
\eeqn 
Another crucial
input is that in the loop channel amplitude a $k$-twisted sector contribution
has to be weighted by the number of fixed points of $\Theta^k$, which are as
well invariant under ${\cal R}$. This factor differs for the given choices 
of the
lattice orientations. For instance for
the six dimensional  {\bf AA} case the loop channel Klein bottle amplitude 
reads 
\beqn
{\cal K} &=&
2c (1-1) \int_0^\infty{\frac{dt}{t^4}\ \left( \frac{\vartheta \left[ 0 
\atop 1/2 \right]^4}{\eta^{12}} \right. }  \non
& & 
\left. \left( \sum_{m\in \mathbb{Z}}{
e^{-4\pi t m^2 /r^2}} \right)^2 \left( \sum_{n\in \mathbb{Z}}{
e^{-3\pi t n^2 r^2}} \right)^2 \right.   \non
& & 
+ \left. \frac{\vartheta \left[ 0 
\atop 1/2 \right]^2}{\eta^6}  
\frac{\vartheta \left[ 1/3 \atop 1/2 \right] 
  \vartheta \left[ -1/3 \atop 1/2 \right]}
  {\vartheta \left[ -1/6 \atop 1/2 \right]
    \vartheta \left[ 1/6 \atop 1/2 \right]} \right. \non
& &  \left. 
+ \frac{\vartheta \left[ 0 
\atop 1/2 \right]^2}{\eta^6} 
\frac{\vartheta \left[ -1/3 
\atop 1/2 \right]\vartheta \left[ 1/3 
\atop 1/2 \right]}{\vartheta \left[ 1/6 
\atop 1/2 \right]\vartheta \left[ -1/6 
\atop 1/2 \right]} \right) ,
\eeqn
the argument being $\exp \left( -4\pi t\right)$. 
After a modular transformation one gets the tree channel amplitude
\beqn
\tilde{\cal K} &=& 
c\frac{32}{3} (1-1) 
\int_0^\infty{dl\ \left( \frac{\vartheta \left[ 1/2 
\atop 0 \right]^4}{\eta^{12}} \right. } \non
& & \left. \left( \sum_{m\in \mathbb{Z}}{
e^{-\pi l m^2 r^2}} \right)^2 \left( \sum_{n\in \mathbb{Z}}{
e^{-4\pi l n^2 /(3r^2)}} \right)^2  \right.   \non
& & + \left. 
3\frac{\vartheta \left[ 1/2 
\atop 0 \right]^2}{\eta^6} \frac{\vartheta \left[ 1/2 
\atop 1/3 \right]\vartheta \left[ 1/2 
\atop -1/3 \right]}{\vartheta \left[ 1/2 
\atop -1/6 \right]\vartheta \left[ 1/2 
\atop 1/6 \right]} \right. \non
& & \left. 
+ 3\frac{\vartheta \left[ 1/2 
\atop 0 \right]^2}{\eta^6} \frac{\vartheta \left[ 1/2 
\atop -1/3 \right]\vartheta \left[ 1/2 
\atop 1/3 \right]}{\vartheta \left[ 1/2 
\atop 1/6 \right]\vartheta \left[ 1/2 
\atop -1/6 \right]} \right) . 
\eeqn
The relative factors $\left( 2\sin (\pi/3 ) \right)^2  
= 3$ between the different contributions to the   
tree channel amplitude come out just as expected to complete the
projector. Switching to the {\bf B} lattice in any complex direction results
in an overall factor of 3, while changing the lattice contributions and
the number of ${\cal R}$ invariant fixed points of $\Theta$. \\

For computing the massless closed string spectra one also needs to distinguish
the various lattice choices very carefully. The untwisted $\Theta$ invariant
states simply have to be symmetrized and antisymmetrized under 
$\Omega{\cal R}$ in the NSNS and the RR sector respectively. 
This contribution to the
spectrum is generic and together with the spectrum of the twisted
sectors has been summarized in table \ref{tableclosed}, where
we left  out the graviton and dilaton
multiplets. 
\TABULAR[h]{|c|c|c|c|}{
\hline
Dim. & Model & untwisted & 
        $\Theta+\Theta^{-1}$  \\ 
& & & twisted \\
\hline 
& {\bf AA} & 3{\mbox H} & 10{\mbox H}+8{\mbox T} \\
$d=6$& {\bf AB} & 3{\mbox H} & 12{\mbox H}+6{\mbox T} \\
& {\bf BB} & 3{\mbox H} & 18{\mbox H} \\ \hline
& {\bf AAA} & 9{\mbox C} & 14{\mbox C}+13{\mbox V} \\
& {\bf AAB} & 9{\mbox C} & 15{\mbox C}+12{\mbox V} \\
\raisebox{1.5ex}[1.5ex]{$d=4$}
& {\bf ABB} & 9{\mbox C} & 18{\mbox C}+9{\mbox V} \\
& {\bf BBB} & 9{\mbox C} & 27{\mbox C} \\
\hline} {\label{tableclosed}Closed string spectra}
The spectrum coming from twisted sectors differs for fixed points
invariant under $\Omega {\cal R}$ and for those fixed points, 
that form pairs under $\Omega {\cal R}$.
The former fixed points contribute two hypermultiplets in six
and a chiral multiplet in four dimensions. The latter fixed points
contribute two hypermultiplets in addition to  two 
tensor multiplets in six dimensions and a chiral plus a vector multiplet 
in four dimensions. 
The sum of the neutral multiplets from the closed string sector finally
matches the sum of the Hodge numbers $h^{1,1}+h^{1,2}$ in all cases.

\subsection{Open string sector: The annulus and the M\"o\-bius strip}

To cancel the tadpoles from the Klein bottle one has to introduce
D($d/2+5$)-branes into the background, which intersect each other on the tori 
at non-trivial angles. They are extended in one dimension on each
two dimensional torus and need to be located  in a $\Theta$ and 
${\cal R}$ invariant way. 
Recalling that open string coordinates with boundary conditions 
\beqn
\Re \left( \frac{\partial}{\partial \sigma} X^i \right) =0 , \quad 
\Im \left( \frac{\partial}{\partial \tau} X^i \right) =0 \nonumber
\eeqn 
at $\sigma=0$ and
\beqn
\Re \left( e^{\pi i w_i} \frac{\partial}{\partial \sigma} X^i \right) =0
 ,\quad 
\Im \left( e^{\pi i w_i} \frac{\partial}{\partial \tau} X^i \right) =0
 \nonumber
\eeqn
at $\sigma=\pi$ have a Fourier decomposition with modings shifted by $w_i$, 
we are led to
consider arrays of 3 kinds of D-branes at relative angle $\pi /3$.  
On each torus $T^2$  one of them is located in the fixed plane of the 
reflection ${\cal R}$ and the remaining ones are obtained by successively  
applying the rotation $\Theta^{1/2}$.     
Such a configuration is shown in Figure \ref{figz3branes}. The type of
lattice is being distinguished, as the number of intersection points of any
two kinds of branes differs. 
\FIGURE[h]{
\makebox[6cm]{
 \epsfxsize=6cm
 \epsfysize=13cm
 \epsfbox{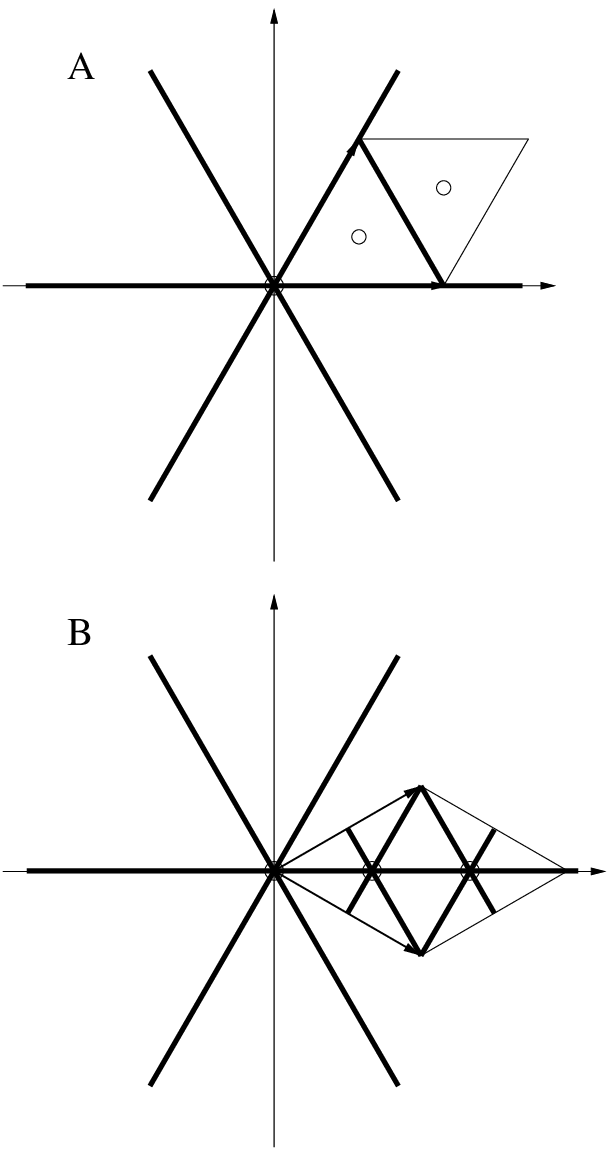}
}
\caption{\label{figz3branes}D-branes on the $SU(3)$ root lattice}}
%
One now needs to compute the annulus 
\beqn
{\cal A}= c & & \int_0^\infty{\frac{dt}{t^{d/2+1}}\  \Tr_{\mbox{\sz open}} 
\left( 
        \frac{1}{2} \frac{1+\Theta +\Theta^2}{3} \right. } \non
& & \left.  \frac{ 1+(-1)^F}{2} e^{-2\pi tL_0} \right) 
\eeqn 
and M\"obius strip amplitudes 
\beqn
{\cal M}= c & & \int_0^\infty{\frac{dt}{t^{d/2+1}}\  \Tr_{\mbox{\sz open}} 
\left( \frac{\Omega{\cal R}}{2} \frac{1+\Theta +\Theta^2}{3} \right. } \non
& & \left. \frac{ 1+(-1)^F}{2} e^{-2\pi tL_0} \right) .
\eeqn
There are only non-vanishing contributions in the traces, 
if the operator in the trace leaves the two D-branes 
$\left( {\rm D}_i,{\rm D}_{i+n} \right)$, the string stretches
between, and its orientation invariant. 
If the moding of the fields in the $\left( {\rm D}_i,{\rm D}_{i+n} \right)$ 
open string sector
is identical to the moding of the fields in the $\Theta^k$ 
twisted closed string
sector, we call such a sector the ``$\Theta^k$ twisted'' open string sector. \\

In order to compute the zero-mode contributions to the 
$\left( {\rm D}_i,{\rm D}_i \right)$
amplitudes, one needs to inspect invariant KK+W
states. The result is sensitive to the type of lattice. 
The additional requirement, that  the M\"obius strip contributions 
have to be  $\Omega{\cal R}$ invariant,  induces a doubling of 
winding numbers as compared to the annulus. 
A very interesting point to notice is an extra multiplicity of some twisted
open string sectors. As can be seen in Figure \ref{figz3branes}, the different
kinds of branes intersect only in the origin of the {\bf A} type unit cell,
while they intersect in all three fixed points of $\Theta$ in the {\bf B} type
unit cell. Thus  the corresponding twisted sectors have  to be weighted 
with extra
multiplicities, leading  to the same extra multiplicity for 
the number of states at each mass level. This turns out to be
necessary  to obtain anomaly free massless spectra. 
Note, that this issue has been related via T-duality to extra factors for
open string sectors
in ordinary $\Omega$ orientifolds with background NS-NS B-field
\cite{hepth9911190}. \\

We present the tree channel annulus and M\"o\-bi\-us strip amplitude
in six dimensions for the {\bf AA} choice of the lattice.
Switching to a {\bf B} type lattice results
in an overall factor of 3 for all amplitudes and has an effect on the
massless spectrum. The annulus reads  
\beqn
\tilde{\cal A} &=& 
c \frac{{\rm M}^2}{6} (1-1) 
\int_0^\infty{dl\ \left( \frac{\vartheta \left[ 1/2 
\atop 0 \right]^4}{\eta^{12}} \right. } \non
& & \left. \left( \sum_{m\in \mathbb{Z}}{
e^{-\pi l m^2 r^2}} \right)^2 \left( \sum_{n\in \mathbb{Z}}{
e^{-4\pi l n^2 /(3r^2)}} \right)^2 \right.  \non
& & \left. +3 \frac{ \vartheta \left[ 1/2 
\atop 0 \right]^2}{\eta^6} 
\frac{\vartheta \left[ 1/2 \atop 1/3 \right]\vartheta \left[ 1/2 \atop -1/3
  \right]}
{\vartheta \left[ 1/2 \atop -1/6 \right]\vartheta \left[ 1/2 \atop 1/6
  \right]} \right. \non
& & \left. + 3 \frac{\vartheta \left[ 1/2 \atop 0 \right]^2}{\eta^6} 
\frac{\vartheta \left[ 1/2 \atop -1/3 \right]\vartheta \left[ 1/2 \atop 1/3
  \right]}
{\vartheta \left[ 1/2 \atop 1/6 \right]\vartheta \left[ 1/2 \atop -1/6
  \right]} \right) . 
\eeqn
with argument $\exp \left( -4\pi l\right)$ and the M\"obius strip gives 
\beqn
\tilde{\cal M} &=& 
-c\frac{8{\rm M}}{3} (1-1) 
\int_0^\infty{dl\ \left( 
\frac{\vartheta \left[ 1/2 \atop 0 \right]^4 \vartheta \left[ 0 \atop 1/2 \right]^4}
{\eta^{12} \vartheta \left[ 0 \atop 0 \right]^4} \right. } \non
& & \left. 
\left( \sum_{m\in \mathbb{Z}}{e^{-4\pi l m^2 r^2}} \right)^2 
\left( \sum_{n\in \mathbb{Z}}{e^{-4\pi l n^2 /(3r^2)}} \right)^2 \right. \non
& & \left. + 3 
\frac{ \vartheta \left[ 1/2 \atop 0 \right]^2  \vartheta \left[ 0 \atop 1/2 \right]^2}
{\eta^6  \vartheta \left[ 0 \atop 0 \right]^2} \right. \\
& & \left. \left( 
\frac{\vartheta \left[ 1/2 \atop 1/3 \right] \vartheta \left[ 0 \atop -1/6
  \right]
\vartheta \left[ 1/2 \atop -1/3 \right] \vartheta \left[ 0 \atop 1/6 \right]}
{\vartheta \left[ 0 \atop 1/3 \right] \vartheta \left[ 1/2 \atop -1/6 \right]
\vartheta \left[ 0 \atop -1/3 \right] \vartheta \left[ 1/2 \atop 1/6 \right]}
\right. \right. \non
& & \left. \left. +  
\frac{\vartheta \left[ 1/2 \atop -1/3 \right] \vartheta \left[ 0 \atop 1/6
  \right]
\vartheta \left[ 1/2 \atop 1/3 \right] \vartheta \left[ 0 \atop -1/6 \right]}
{\vartheta \left[ 0 \atop -1/3 \right] \vartheta \left[ 1/2 \atop 1/6 \right]
\vartheta \left[ 0 \atop 1/3 \right] \vartheta \left[ 1/2 \atop -1/6 \right]} 
\right) \right) . \nonumber
\eeqn
with argument $\exp \left( -8\pi l\right)$.
One obtains only a  single overall tadpole cancellation condition, which reads
\beqn \nonumber
\frac{1}{6} \left({\rm M}^2-16{\rm M}+64 \right) = \frac{1}{6} 
\left( {\rm M}-8 \right)^2 =0 .
\eeqn
It implies the presence of ${\rm M}=8$ D7-branes of each kind. The analogous
condition in four dimensions requires ${\rm M}=4$ D6-branes. The tadpoles are
related to contributions of 8- and 7-forms respectively. 
Thus,  we get an $SO(8)$ gauge
group in six dimensions and $SO(4)$ in four. The rank reduction with respect
to the $SO(32)$ in ten dimensions can be related to the rank reduction in
ordinary $\Omega$ orientifolds with background B-field via T-duality.
As was pointed out in \cite{hepth9912218}
it is possible via discrete Wilson-lines
to get the gauge groups $Sp(8)$ and $Sp(4)$, respectively.  \\

The massless spectrum in the untwisted open string sector is again generic. In
six dimensions there is a hypermultiplet, in four dimensions three chiral
multiplets in addition to the gauge vectors, all in the antisymmetric
representation. In the twisted sectors one needs to count the extra
multiplicity of states deriving from the number of intersection points, each
giving rise to a hypermultiplet in the antisymmetric in six and a chiral
multiplet in the symmetric representation in four dimensions. The open string
spectra are  collected in table \ref{tableopenz3}.
\TABULAR[h]{|c|c|c|c|}{
\hline 
 Dim. & Model & $\left( {\rm D}_{i},{\rm D}_{i} \right)$ & 
        $\left( {\rm D}_{i},{\rm D}_{i+1} \right)$  \\
\hline 
 & ${\bf AA}$ & $1{\rm H\ in\ } ({\bf 28})$ & $1{\rm H\ in\ } ({\bf 28})$ \\
$d=6$  
 & ${\bf AB}$ & $1{\rm H\ in\ } ({\bf 28})$ & $3{\rm H\ in\ } ({\bf 28})$ \\
 & ${\bf AA}$ & $1{\rm H\ in\ } ({\bf 28})$ & $9{\rm H\ in\ } ({\bf 28})$ \\
\hline
 
 & ${\bf AAA}$ & $3{\rm C\ in\ } ({\bf 6})$ & $1{\rm C\ in\ } ({\bf 10})$ \\
 & ${\bf AAB}$ & $3{\rm C\ in\ } ({\bf 6})$ & $3{\rm C\ in\ } ({\bf 10})$ \\
\raisebox{1.5ex}[1.5ex]{$d=4$}
 & ${\bf ABB}$ & $3{\rm C\ in\ } ({\bf 6})$ & $9{\rm C\ in\ } ({\bf 10})$ \\
 & ${\bf BBB}$ & $3{\rm C\ in\ } ({\bf 6})$ & $27{\rm C\ in\ } ({\bf 10})$ \\
\hline}{\label{tableopenz3} Open string spectra}
All the six dimensional models satisfy the cancellation condition of the
irreducible anomaly
\beqn
n_{\rm H} -n_{\rm V} +29 n_{\rm T}=273
\eeqn
while the four dimensional models are free of any gauge anomaly anyway.

\section{Results for $\mathbb{Z}_4$, $\mathbb{Z}_6$ and $\mathbb{Z}_6^\prime$} 

In this section we shall only briefly explain some further properties of 
the other
models which we have  computed so far. The lattice and brane
configurations, the explicit formulas for all the amplitudes to be calculated
as well as all their massless spectra have been presented in
\cite{hepth9908130,hepth9912204}. \\

The world-sheet consistency condition is more restrictive than for the
$\mathbb{Z}_3$, where all combinations of lattices were allowed. For all other
examples we find that the relative orientation is fixed between two of the
complex tori, while, in four dimensions, the third can be chosen freely of
{\bf A} or {\bf B} type. With this freedom we obtain another set of eight
inequivalent models from the three orbifold groups. \\
In the open string sector one again has to introduce sets of branes at $\pi
v_i$ relative angles, where their respective number has to be determined by
the tadpole cancellation conditions.
Since there  always exist two types of branes
which are not mapped upon each other by the orbifold, the gauge group will
have a product structure with two identical factors. The spectra, however,
need not be symmetric under exchanging the two factors. 
As all the orbifold groups contain an element $\Theta^{N/2}$ of order
two, there is also a contribution to the $\mathbb{Z}_2$ twisted sector in
the tree channel. This generically imposes the second tadpole cancellation
condition  
\beqn
\tr \left( \gamma_{N/2}^{(i)} \right)=0 
\eeqn 
for the action of $\Theta^{N/2}$ on the Chan-Paton factors of each type 
$i$ of branes. The two
conditions together then exactly reproduce the situation of the standard
six dimensional $\mathbb{Z}_2$ orientifold discussed by Bianchi and
Sagnotti   
in \cite{sagbi} and later by Gimon and Polchinski in \cite{hepth9601038}. 
Their solution implies that the
orthogonal $SO({\rm M})$ gauge group is broken to its unitary $U({\rm M}/2)$
subgroup. For instance, in four dimensions for 
the $\mathbb{Z}_4$ orbifold we find two possible solutions $U(8)^2$ and
$U(4)^2$, while for the various $\mathbb{Z}_6$ models we always get
$U(2)^2$. This is in accord with the rank reduction expected from T-duality.\\
 
The determination of the massless spectra is in general more involved than for
the $\mathbb{Z}_3$. In the twisted sectors of the 
closed string spectrum one has to keep track of the transformation properties
of the various fixed points under the orbifold group generator 
$\Theta$ and $\Omega{\cal R}$ in order to find the correct symmetrization
prescription. Also one needs to distinguish with respect to the action of
$\Theta^{N/2}$. In the open string spectrum the same is true for the actions
on the intersection points, which do not simply provide all the same
states anymore. In four dimensions there are certain subtle phase factors to be
regarded, which sometimes require extra signs in the loop channel amplitude in
order to complete the tree channel projector. Finally, for the most
complicated case, the $\mathbb{Z}_6^\prime$, one even needs to distinguish
different types of branes with different effective gauge theories on them. 
The results display spectra which are anomaly free in six dimensions and
non-chiral in four.

\acknowledgments{ 
We would like to thank C. Angelantonj, who participated in closely related
work, A. Kumar, who was also involved in an early stage, 
as well as M. Gaberdiel and D. L\"ust for encouraging discussions and helpful remarks.  

\bibliographystyle{JHEP}
\bibliography{proceedings}

\providecommand{\href}[2]{#2}\begingroup\raggedright\begin{thebibliography}{10}

\bibitem{hepth9510169}
J.~Polchinski and E.~Witten, {\it Evidence for heterotic - type {I} string
  duality},  {\em Nucl. Phys.} {\bf B460} (1996) 525--540,
  [\href{http://xxx.lanl.gov/abs/hep-th/9510169}{{\tt hep-th/9510169}}].

\bibitem{Aldazabal:1998mr}
G.~Aldazabal, A.~Font, L.~E. Ibanez, and G.~Violero, {\it {D} = 4, {N} = 1,
  type {IIB} orientifolds},  {\em Nucl. Phys.} {\bf B536} (1998) 29,
  [\href{http://xxx.lanl.gov/abs/hep-th/9804026}{{\tt hep-th/9804026}}].

\bibitem{hepth9804092}
Z.~Kakushadze, G.~Shiu, and S.~H.~H. Tye, {\it Type {IIB} orientifolds,
  {F}-theory, type {I} strings on orbifolds and type {I} heterotic duality},
  {\em Nucl. Phys.} {\bf B533} (1998) 25,
  [\href{http://xxx.lanl.gov/abs/hep-th/9804092}{{\tt hep-th/9804092}}].

\bibitem{hepth9908130}
R.~Blumenhagen, L.~G{\"o}rlich, and B.~Kors, {\it Supersymmetric orientifolds
  in 6{D} with {D}-branes at angles},
  \href{http://xxx.lanl.gov/abs/hep-th/9908130}{{\tt hep-th/9908130}}.

\bibitem{hepth9912204}
R.~Blumenhagen, L.~G{\"o}rlich, and B.~Kors, {\it Supersymmetric 4{D}
  orientifolds of type {IIA} with {D}6-branes at angles},
  \href{http://xxx.lanl.gov/abs/hep-th/9912204}{{\tt hep-th/9912204}}.

\bibitem{hepth9912218}
G.~Pradisi, {\it Type {I} vacua from diagonal {Z}(3)-orbifolds},
  \href{http://xxx.lanl.gov/abs/hep-th/9912218}{{\tt hep-th/9912218}}.

\bibitem{hepth9911190}
C.~Angelantonj and R.~Blumenhagen, {\it Discrete deformations in type {I}
  vacua},  \href{http://xxx.lanl.gov/abs/hep-th/9911190}{{\tt hep-th/9911190}}.

\bibitem{DHVW1985}
L.~Dixon, J.~A. Harvey, C.~Vafa, and E.~Witten, {\it Strings on orbifolds},
  {\em Nucl. Phys.} {\bf B261} (1985) 678--686.

\bibitem{DHVW1986}
L.~Dixon, J.~A. Harvey, C.~Vafa, and E.~Witten, {\it Strings on orbifolds. 2},
  {\em Nucl. Phys.} {\bf B274} (1986) 285--314.

\bibitem{hepth9904007}
Z.~Kakushadze, {\it Non-perturbative orientifolds},  {\em Phys. Lett.} {\bf
  B455} (1999) 120, [\href{http://xxx.lanl.gov/abs/hep-th/9904007}{{\tt
  hep-th/9904007}}].

\bibitem{sagbi}
M.~Bianchi and A.~Sagnotti, {\it Twist symmetry and open string wilson lines},
  {\em Nucl. Phys.} {\bf B361} (1991) 519.

\bibitem{hepth9601038}
E.~G. Gimon and J.~Polchinski, {\it Consistency conditions for orientifolds and
  {D}-manifolds},  {\em Phys. Rev.} {\bf D54} (1996) 1667--1676,
  [\href{http://xxx.lanl.gov/abs/hep-th/9601038}{{\tt hep-th/9601038}}].

\end{thebibliography}\endgroup
\end{document}